\documentstyle[preprint,aps]{revtex}

\begin{document}

\draft

\title{Monte Carlo with Absorbing Markov Chains:\\
Fast Local Algorithms for Slow Dynamics}

\author{M.~A. Novotny}

\address{Supercomputer Computations Research Institute,
Florida State University,
Tallahassee, FL 32306-4052}

\date{\today}

\maketitle

\begin{abstract}
A class of Monte Carlo algorithms which incorporate absorbing Markov
chains is presented.  In a particular limit,
the lowest-order of these algorithms
reduces to the $n$-fold way algorithm.
These algorithms are applied to study the escape from
the metastable state in the two-dimensional square-lattice
nearest-neighbor Ising ferromagnet in an unfavorable
applied field, and the agreement with
theoretical predictions is very good.  It is demonstrated that the
higher-order algorithms can be many orders of magnitude faster than either the
traditional Monte Carlo or $n$-fold way algorithms.
\end{abstract}

\vskip 0.5 true cm
\pacs{02.70.Lq, 05.50.+q, 64.60.My, 75.40.Mg}
%02.70.Lq Monte Carlo and Statistical Methods
%05.50.+q Lattice Theory and Statistics, Ising Problems
%64.60.My Metastability
%75.40.Mg Numerical Simulation Studies
%
\narrowtext

%\section{INTRODUCTION}
%\typeout{INTRODUCTION}
%\subsection{INTRODUCTION MORE}
%\typeout{Subsection A: INTRODUCTION MORE}

Monte Carlo (MC) methods\cite{re:Binder1} have become indispensible tools
for nonperturbative calculations in numerous fields, including
materials science, high-energy physics, chemistry, biology,
engineering, and economics.  These methods are used for two fundamentally
different purposes: to calculate time-independent quantities (statics) and
to simulate time series (dynamics).  In the former case, the slow
relaxation observed, e.g.\ near phase transitions (critical slowing down)
and at low temperatures, is merely a nuisance that has been overcome
by a number of new MC algorithms, including cluster algorithms\cite{re:clust},
vertex algorithms\cite{re:vertx}, multicanonical algorithms\cite{re:Berg},
and hybrid MC algorithms\cite{re:hybrid}.
%Monte Carlo (MC) methods\cite{re:Binder1} are used to calculate
%quantities in a nonperturbative fashion in fields of science
%including, among others,
%critical phenomena, materials science, biology, and high-energy physics.
%There has recently been significant progress in new MC
%algorithms which can overcome the slow relaxation near phase transitions
%(critical slowing down) and at low temperatures.  These algorithms include
%cluster algorithms\cite{re:clust}, vertex algorithms\cite{re:vertx},
%multicanonical algorithms\cite{re:Berg}, and hybrid MC
%algorithms\cite{re:hybrid}.
Such algorithms can be many orders of
magnitude faster than standard MC methods.  However, they all replace the
standard MC dynamic with a different dynamic.  Consequently, although such
algorithms may be very efficient in the
calculation of static quantities, information about the
kinetics of the original MC dynamic cannot be obtained.
There are many instances where the kinetics, rather than just the statics,
is of physical importance.  Recently, MC methods using
constrained cluster-flipping algorithms have been proposed in order to
obtain information about the long-wavelength kinetics of a
system\cite{re:BarkMark}.  However, in such methods the local dynamic
is modified, and universality arguments must be made to relate
the results to the dynamic of the original system.

In this Letter, a different class of
accelerated MC algorithms is presented that does
not change the original MC dynamic, and can therefore be used to
simulate time series with very large separations in time scales.
These algorithms incorporate into the standard MC algorithm
absorbing Markov chains (AMC).  The acronym MCAMC stands for
Monte Carlo with absorbing Markov chains.
With some additional approximations, the lowest-order MCAMC algorithm
corresponds to the $n$-fold way algorithm \cite{re:nfold}.
We demonstrate that MCAMC algorithms can be many orders of
magnitude faster than either the standard MC or $n$-fold way algorithms.

In a Markov process the probability of going from the
current state of the system to another state depends only on these two
states, not on the history of how the current state was reached.
Standard MC algorithms are typical examples of Markov processes.
A Markov process with a finite state space and a discrete
time step is a Markov chain.  The Markov chain is often written as a
Markov matrix, ${\bf M}$, the elements of which are the transition
probabilities between the states.  The time evolution of the probability
distribution vector ${\vec v}^{\rm T}$ is then given by
${\vec v}^{\rm T}(m+1) = {\vec v}^{\rm T}(m){\bf M}$.
An AMC is one in which at least one
state has the property that transitions out of that state are forbidden.

The Markov matrix associated with an AMC with $q$ absorbing states
and $s$ transient states is given by\cite{re:MCbook}
\begin{equation}
\label{eq:AMC0}
{\bf M} = \pmatrix{{\bf I}_{q\times q} & {\bf 0}_{q\times s} \cr
                   {\bf R}_{s\times q} & {\bf T}_{s\times s} \cr} ,
\end{equation}
where ${\bf I}$ is the identity matrix,
${\bf 0}$ is the zero matrix, and the subscripts show the
size of each submatrix.  The elements of the transient submatrix, ${\bf T}$,
are the transition probabilities between the $s$ transient
states of ${\bf M}$.
The elements of the submatrix ${\bf R}$ are the transition probabilities
from each of the $s$ transient states to the $q$ absorbing states.
Starting from an initial vector $\vec v_I^{\rm T}$, the probability of
still being in any of the $s$ transient states after $m$ time steps is given
by $\vec v_I^{\rm T} {\bf T}^m \vec e$, where $\vec e$ is the vector with
all elements equal to~1.  Although $\vec v_I^{\rm T}$ could be any
normalized probability distribution over the transient states, in this
work we will take $\vec v_I^{\rm T}$
to represent one particular transient state.
%Using a random number $r$, uniformly distributed
%over the interval $(0,1]$, one can obtain the time $m$ for exiting to one
%of the $q$ absorbing states from the solution of the equation
It is easy to show that one can obtain the time $m$ for exiting to one of
the $q$ absorbing states from the solution of the equation
\begin{equation}
\label{eq:AMC1}
{\vec v}_I^{\rm T} {\bf T}^m {\vec e} < r \le
{\vec v}_I^{\rm T} {\bf T}^{m-1} {\vec e} \;,
\end{equation}
where $r$ is a random number uniformly distributed on the interval $(0,1]$.
Note that the stochastic variable $m$ is independent of which of the
$q$ states the AMC ends up in.
The elements of the vector of $q$ unnormalized probabilities
\begin{equation}
\label{eq:AMC2}
\vec {\cal Q}^{\rm T} =
{\vec v}_I^{\rm T} {\bf T}^{m-1}_{s\times s} {\bf R}_{s\times q}
\end{equation}
give the probability of exiting to each corresponding state, given that the
system has exited from the transient subspace in $m$ time steps.
After normalization, this probability distribution can be used to pick a
particular state into which the AMC exits.
This is done by using another uniformly-distributed random number.
Eqs.~(\ref{eq:AMC1}) and (\ref{eq:AMC2}) are the only equations needed
to include absorbing Markov chains within a Monte Carlo simulation.
In practice, it is computationally sound policy to
obtain all the eigenvalues and eigenvectors of ${\bf T}$
and utilize the spectral decomposition of ${\bf T}$
within the AMC portion of the algorithm to
numerically solve Eqs.~(\ref{eq:AMC1}) and (\ref{eq:AMC2}).

The MCAMC algorithm has the following steps:  0)~The system is in an
initial state represented by ${\vec v}_I^{\rm T}$.  1)~Write an
$s$$\times$$s$ transient matrix that includes at least the initial
configuration, but ideally will include other states (which may
represent many configurations of the system that are related by symmetry).
The additional states included as transient states should be those the
current configuration has the largest probability of exiting to.
2)~All configurations that the transient states can exit to in one
time step must be included in the absorbing states.
3)~Use Eq.~(\ref{eq:AMC1}) to find the time spent in the transient
state subspace, and Eq.~(\ref{eq:AMC2}) to find a new configuration for the
system.  4).~Iterate the procedure using the new configuration as the
next initial configuration.  Thus, once the system has exited to one of the
$q$ absorbing states, this decides the initial configuration
of the system for a different AMC with different transient
and absorbing subspaces, and the process is iterated.

To illustrate the MCAMC algorithm, we apply it to the square-lattice
nearest-neighbor Ising ferromagnet in a magnetic field.
The Hamiltonian is given by
%\begin{equation}
%\label{eq:Ham}
${\cal H} = -J \sum_{\langle i,j\rangle} s_i s_j -H\sum_i s_i$,
%\end{equation}
where the spins $s_i$$=$$\pm 1$.  The sums run over
all nearest-neighbor pairs and over all $N$$=$$L^2$ sites, respectively.
Periodic boundary conditions are used.
The isotropic two-body coupling constant is given by $J$$>$$0$ and the
magnetic field by $H$.
To study the decay of a metastable phase,
we apply a negative magnetic field at a temperature below the
critical temperature $T_c$, and
start with the configuration of all spins $+1$ (which we call the
C+ configuration).
Standard droplet theory (for a recent review on metastability see
Ref.\ \cite{re:PER3}) shows that in order for the metastable phase
C+ to decay, one or more critical droplets must nucleate via the
random superposition of microscopic fluctuations.  Since the critical
droplet is of a certain size, the average time before its creation, and
thus the metastable lifetime, is much longer than the microscopic
timescales.
Although any local dynamic can be used within
the MCAMC framework, in this letter only Metropolis updates
\cite{re:Binder1} are performed.
We measure the number of Monte Carlo steps per spin (MCSS) until a
configuration is reached which has an equal number of $+1$ and $-1$ spins.
The average lifetime, $\tau$, of the metastable state is found by
averaging over a number of starts from the C+ configuration.

A standard MC algorithm randomly chooses a spin, and then decides
whether or not to flip it.
Each spin in the $L$$\times$$L$ lattice, with periodic boundary conditions,
is in one of $10$ possible energy classes
which are determined by how the spin is oriented with
respect to the applied field and the nearest-neighbor
spins \cite{re:nfold}.
The number of spins in class $i$ is $n_i$, and the
probability of flipping a spin in class $i$ once it has been chosen is $p(i)$.

The $s$$=$$1$ MCAMC algorithm is defined to have a single
transient state, which is the current
state of the spin configuration.  When spins belonging to each
class are grouped together the submatrix
${\bf R}_{1\times 10}=\big (n_1p(1), \cdots, n_{10}p(10)\big )/N$.
Define $Q_0$$=$$0$ and $Q_j$$=$$\sum_{i=1}^j n_i p(i)$.
Since ${\bf M}$ is a
Markov matrix, its row-sums are unity, so the
transient matrix is ${\bf T}_{1\times 1}$$=$$1$$-$$Q_{10}/N$.
The time increment to flip a spin is found from
Eq.~(\ref{eq:AMC1}) to be
$m$$>$$\ln(r)$$/$$\ln(1$$-$$Q_{10}/N)$$\ge$$m$$-$$1$.
If one relaxes the constraint that $m$ is an integer, and expands to lowest
order for small $Q_{10}/N$, this equation becomes
$\widehat m$$=$$-N\ln(r)/Q_{10}$,
and one has the standard $n$-fold way algorithm \cite{re:nfold}.
In both the $s$$=$$1$ MCAMC and the $n$-fold way algorithm,
another random number
is used to choose which of the $10$ classes to pick via
$Q_{j-1}$$<$$rQ_{10}$$\le$$Q_{j}$,
and a spin from class $j$ is randomly picked and flipped.
As shown in Fig.\ \ref{fig1}, at low temperatures the $s$$=$$1$ MCAMC algorithm
can be many orders of magnitude faster than the standard MC algorithm.
At $J/T$$=$$3$ the speed improves by a factor of about $10^7$.

At low temperatures in the $s$$=$$1$ MCAMC algorithm, once a spin has
flipped from the C+ configuration,
it is extremely probable that in the next iteration this spin will be chosen
again to flip, and one returns to the C+ configuration.
This problem can be addressed by going to
the $s$$=$$2$ MCAMC algorithm.  Whenever
the starting configuration is the C+ configuration or a configuration
with only one spin overturned from the C+ configuration,
one includes these two states in the transient matrix ${\bf T}$.
The normal $s$$=$$1$ MCAMC algorithm is used if the spin configuration
is other than one of the transient states.
As shown in Fig.\ \ref{fig1}, for $J/T$$=$$3$ the speed
improves by a factor of about
$10^2$ over the $n$-fold way algorithm.

One can continue to increase the number of states in ${\bf T}$.
Figure \ref{fig1} also includes results from $s$$=$$3$ MCAMC, where
the three states in ${\bf T}$ were C+, C+ with one overturned spin,
and C+ with two nearest-neighbor overturned spins.
At $J/T=3$ this $s$$=$$3$ MCAMC algorithm improves the speed
by a factor of about $10^2$ compared to the $s$$=$$2$ MCAMC algorithm.

The MCAMC algorithm offers a further advantage, in that the statistical
error in the average lifetime
$\tau$ is smaller than from standard MC.
The average lifetime of an AMC
that starts from state $k$ is given by $\vec v_k^{\rm T} {\bf N} \vec e$,
where the fundamental matrix is defined by
${\bf N}$$=$$({\bf I}$$-$${\bf T})^{-1}$ \cite{re:MCbook}.
For $s$$>$$1$, the initial configuration in which the AMC starts is
C+ only once, since it must re-enter C+ from a configuration
C+ with one overturned spin.
Since the contribution to the lifetime that comes from
the escape from C+ is exact, this reduces the total error in $\tau$.

It has been shown that at low enough temperatures $\tau$ is
related to the height of the lowest energy barrier which must be reached
by flipping one spin at a time starting from C+ \cite{re:lowT1}.
In the limit of low temperature and large $L$,
the discreteness of the lattice gives an important contribution
to $\tau$.  Theorem~3 of Ref.~\cite{re:lowT1} states that then
\begin{equation}
\label{eq:lowT}
k_B T \ln(L^2\tau) = \Gamma(H,J) = 8J\ell_c -2|H|(\ell_c^2-\ell_c+1)
\end{equation}
where $\ell_c$$=$$\lceil 2J/|H|\rceil$
and the notation $\lceil x\rceil$ denotes
the smallest integer larger than $x$.
This result, which is restricted to $2J$$/$$|H|$
not an integer
and to $|H|$$<$$4J$, is shown in Fig.\ \ref{fig1} as a solid
line.  It is in qualitative agreement with the measured values of $\tau$.
A detailed comparison with predictions from Ref.~\cite{re:lowT1} will be
published elsewhere \cite{re:NovUn}.

Using similar reasoning, it is possible to estimate at low temperatures
the temperature dependence of the different MCAMC algorithms.
By assuming that most of the CPU time is spent in the $s$$=$$1$ portion
of the algorithm, at low enough temperatures the CPU time should be
proportional to
$\exp\{[\Gamma(H,J)$$-$$\Gamma_0(H,J)]$$/$$k_B T\}$.  Here
$\Gamma_0$ is $2J$ times the surface area minus $2|H|$ times the volume of
the largest compact lattice animal included in ${\bf T}$.  The dashed lines in
Fig.\ \ref{fig1} have these slopes and are drawn to go through the data point
at the lowest temperature available for a particular algorithm.

Figure \ref{fig2} shows values for $\tau$ as a function of $|H|$ for two
low temperatures.  Note how well Eq.~(\ref{eq:lowT}) fits the results.
Eq.~(\ref{eq:lowT}) is only valid if the nucleation of a single droplet
is responsible for flipping the lattice into the stable phase.
A reasonable estimate for the crossover field
(which has been called the dynamic spinodal field \cite{re:PER1})
out of the single-droplet regime
is $H_{1/2}$, the field at which
the standard deviation of the lifetime is $\tau$$/$$2$ \cite{re:PER1}.
Fig.\ \ref{fig2} shows that this is a reasonable estimate
for the point at which
the results deviate from Eq.~(\ref{eq:lowT}) at strong fields.

At higher temperatures, the discreteness of the lattice becomes
less important, and $\tau$ should be given by the continuum
droplet-theory prediction \cite{re:drop0,re:drop1,re:drop2,re:drop3}
\begin{equation}
\label{eq:droplet}
\ln(\tau) = \Xi(T)/|H| -(b+c)\ln|H| + \ln\big(A(T)\big).
\end{equation}
Here $\Xi(T)$ is related to the equilibrium zero-field surface tension,
the equilibrium droplet shape \cite{re:Zia}, and the
spontaneous magnetization.  $A(T)$ is a non-universal function.
Field-theoretical and numerical calculations give
$b$$=$$1$ \cite{re:drop1,re:drop3,re:PER2}.
For dynamics described by a Fokker-Planck equation
$c$$=$$2$ \cite{re:drop2,re:drop3},
and a recent MC study is consistent with $b$$+$$c$$=$$3$ \cite{re:PER1}.

Differentiating Eqs.~(\ref{eq:lowT}) and (\ref{eq:droplet}) with respect to
$|H|^{-1}$ allows for direct comparisons with both the discrete droplet
and continuum droplet predictions.  This is shown in Fig.\ \ref{fig3}.
For $T$$=$$0.4$$J$
a crossover is observed at weak $|H|$ to the continuum droplet
prediction using $b$$+$$c$$=$$3$ and the exact zero-field
value for $\Xi(T)$ \cite{re:Zia}.  At weak $|H|$ Fig.\ \ref{fig3} shows
that the $T$$=$$0.4$$J$ data are
consistent with
the theoretical predictions of Eq.~(\ref{eq:droplet}) with the
theoretical values for $\Xi(T)$ and $b$$+$$c$.

MCAMC algorithms can be utilized to study other interesting low-temperature
effects in lattice-gas types of models.
An example is the recent prediction that at low
enough temperatures in the anisotropic Ising model nucleation occurs
through square droplets, rather than through rectangular equilibrium
Wulff droplets \cite{re:lowTaniso}.

Additional applications of MCAMC algorithms may include studies of
equilibrium properties of systems where the $n$-fold way algorithm has
been utilized.  These include the anisotropic Ising model \cite{re:Landau}
(which is related via a Trotter-Suzuki decomposition to a quantum
model in one lower dimension \cite{re:Suzuki}),
and simulated annealing approaches
to minimization \cite{re:SimAnn}.

However, the real strength of MCAMC algorithms lies in the study of slow
dynamics in models with a limited number of local states.
MCAMC algorithms will be particularly useful when there exists
a small number of
spins which are rapidly fluctuating, and major rearrangements
of spins occur very infrequently.  Examples would be
the kinetics of spin-glass models \cite{re:LATER},
where the spins that fluctuate rapidly are the
ones that find themselves in a local environment with zero
energy difference between the possible spin orientations.
Another example is phase-ordering kinetics in which
a few particles undergo rapid random walks on
long flat portions of interfaces, while the interfaces move
extremely slowly.  Similar reasoning also applies to
simulations of molecular-beam epitaxial  growth,
where the $n$-fold way algorithm
has been rediscovered at least twice \cite{re:Gilmer,re:Maksym}.
In all such cases, using higher-order
MCAMC algorithms should substantially decrease the CPU time required
to obtain the static and dynamic information about the system without making
any approximations to the dynamics.

\acknowledgments

The author wishes to thank J.~Lee and P.~A.~Rikvold for useful discussions
and suggestions.
This research was supported in part by the Florida State University
Supercomputer Computations Research Institute, which is partially funded
through contract \# DE-FC05-85ER25000 by the U.S. Department of Energy.

\vskip 0.0 true in
\eject

%--------------------------------------------------------------

\begin{figure}
\caption{
The average CPU time for escape from the metastable state is shown as a
function of the inverse temperature.  This is for a $24$$\times$$24$ lattice
in a field of $|H|/J$$=$$0.75$.
For the standard Monte Carlo
algorithm ($\times$) the CPU time is directly proportional to the average
lifetime $\tau$ of the metastable state, and $\tau$
(in units of Monte Carlo steps per spin) is plotted on
the right-hand axis.  All values of $\tau$ were calculated by
averaging over $10^3$ escapes from the metastable state.
The symbol $\times$, with error estimates, for $J/T$$\ge$$1$
is calculated from the values of $\tau$ from the MCAMC algorithms, while
for $J/T$$<$$1$ the CPU time for a standard Monte Carlo algorithm is plotted.
The CPU time from the $n$-fold way algorithm, which corresponds to
$s$$=$$1$ MCAMC, is given by the symbol $\Diamond$.
The timings for $s$$=$$2$ and $s$$=$$3$ MCAMC algorithms are given by
the symbols + and $\Box$, respectively.
The vertical arrow marks the exact critical temperature.
The solid line is the low-temperature
discrete droplet result \protect\cite{re:lowT1}
for $\tau$ with $\ell_c$$=$$3$ from
Eq.~(\protect\ref{eq:lowT}).
The dashed lines are estimates, described in the text, for the CPU times for
the MCAMC algorithms.
The horizontal arrows indicate that the solid line and symbols $\times$
are related to both vertical axes, whereas the other points and lines
are only for the left-hand axis.
The curvature near $J$$/$$T$$\approx$$1$ is near the value where
$H_{1/2}$$=$$3$$J$$/$$4$ for this system size.
Note the spectacular increases in speed with the MCAMC algorithms.}
\label{fig1}
\end{figure}

\begin{figure}
\caption{
The average lifetime $\tau$, as a function of $|H|$, for escape from the
metastable state is shown for a $24$$\times$$24$ lattice with
$T/J$$=$$0.2$ ($T/T_c$$\approx$$0.088$) ($\times$)
and $T/J$$=$$0.4$ ($\bigcirc$).
Error estimates are plotted for each point.
The solid lines are the low-temperature discrete droplet estimates for
$\tau$ from Eq.~(\protect\ref{eq:lowT}).
The diagrams show the nucleating droplet, reading from the
strongest fields with $\ell_c$ of 1, 2, and 3.  The dashed lines show the
boundary between the various $\ell_c$ regions.
The vertical arrows mark the cross-over
field $H_{1/2}$ \protect\cite{re:PER1},
which separates the single-droplet
regime ($|H|$$<$$H_{1/2}$) and the multidroplet regime ($H_{1/2}$$<$$|H|$)
for the two temperatures for this lattice size.}
\label{fig2}
\end{figure}

\begin{figure}
\caption{
The temperature times the derivative of the logarithm of the
average lifetime with respect to $|H|^{-1}$ is shown as a function of $|H|$.
This is for a $24$$\times$$24$ lattice with
$T$$=$$0.2$$J$ ($\times$) and $T$$=$$0.4$$J$ ($\bigcirc$).
The solid curves are from the low-temperature
predictions \protect\cite{re:lowT1} from Eq.~(\protect\ref{eq:lowT}).
Only results in the single droplet regime, $|H|$$\le$$H_{1/2}$, are
shown.
The dot-dashed and dashed horizontal lines correspond to the
exact\protect\cite{re:Zia} zero-field value for
$\Xi(T)/T$ for $T$$=$$0.2$$J$ and $T$$=$$0.4$$J$,
respectively.  The dashed inclined straight line includes the
theoretical value $b$$+$$c$$=$$3$ in
Eq.~(\protect\ref{eq:droplet}) for $T$$=$$0.4$$J$.}
\label{fig3}
\end{figure}

\end{document}